\documentclass[aps,prb,twocolumn,superscriptaddress,showpacs]{revtex4-1}
\usepackage{longtable}
\usepackage{graphicx}% Include figure files
\usepackage{subfigure}
\usepackage{dcolumn}% Align table columns on decimal point
\usepackage{bm}% bold math
\usepackage{amsmath}
\usepackage[psamsfonts]{amssymb}
\usepackage{lineno}

\newcommand{\YBCO}{\text{YBa$_{2}$Cu$_{3}$O$_{7-\delta}$}~}
\newcommand{\CaYBCO}{\text{Y$_{1-x}$Ca$_{x}$Ba$_{2}$Cu$_{3}$O$_{7-\delta}$}~}
\newcommand{\LAO}{\text{LaAlO$_{3}$}~}

\newcommand{\Tc}{\text{$T_{c}$}~}
\newcommand{\cm}{\text{cm$^{-1}$}~}
\newcommand{\sigr}{$\sigma_{1}(\omega,T)$~}
\newcommand{\sigi}{$\sigma_{2}(\omega,T)$~}
\newcommand{\wsig}{$\omega {{\sigma }_{2}}(\omega ,T)$}
\newcommand{\etal}{\emph{et al.}}
\newcommand{\dwave}{$d_{{{x}^{2}}-{{y}^{2}}}$-wave~}

\begin{document}

%\linenumbers

%Title of paper
\title{Evidence of a subenergy gap in the overdoped regime of \CaYBCO thin films from THz Spectroscopy}

\author{N. Bachar}
\email[]{nimib@ariel.ac.il}
\altaffiliation{Department of Applied Physics, Hebrew University, Jerusalem, Israel (Previous affiliation)}
\affiliation{Laboratory for Superconductivity and Optical Spectroscopy, Ariel University Center of Samaria, Ariel, 40700, Israel}
\affiliation{The Raymond and Beverly Sackler School of Physics $\&$ Astronomy, Tel Aviv University, Tel Aviv, 69978, Israel}

\author{E. Farber}
\affiliation{Laboratory for Superconductivity and Optical Spectroscopy, Ariel University Center of Samaria, Ariel, 40700, Israel}

\author{E. Zhukova}
\affiliation{Submillimeter Spectroscopy Department, Prokhorov General Physics Institute, Russian Academy of Sciences, Moscow, Russia}

\author{B. Gorshunov}
\affiliation{Submillimeter Spectroscopy Department, Prokhorov General Physics Institute, Russian Academy of Sciences, Moscow, Russia}

\author{M. Roth}
\affiliation{Department of Applied Physics, Hebrew University, Jerusalem, Israel}

\date{\today}

\begin{abstract}

We measured the terahertz (THz) complex conductivity of Ca doped \YBCO thin films in the frequency range of 0.1 to 3~THz (3 to 100~\cm) and at a temperature range of 20 to 300~K. The films were measured using both time domain and frequency domain THz methods. We showed evidence for the existence of a sub-gap in overdoped \CaYBCO samples doped with 5\% and 10\% Ca. Evidence for the opening of this sub-gap appears as a sharp decrease in the spectrum of the real part of conductivity at frequencies equivalent to a gap energy of 1~meV and is more prominent with increased doping. This decrease in conductivity can be explained by using \dwave pairing symmetry with an imaginary part of $is$ or $i{{d}_{xy}}$ which suggests node removal.

\end{abstract}



\maketitle

Symmetry of the order parameter in high-\Tc cuprates superconductors and the existence of nodal lines along the diagonal of the Brillouin zone, the $\Gamma$X direction, has been and is still one of the theoretical and experimental questions to be addressed~\cite{Deutscher2005}. There is a degree of consensus that the pairing symmetry consists of a dominant \dwave where there is an extra imaginary component which is doping and magnetic dependent. This extra component is still puzzling and leads us to important questions that remain to be resolved. One of these questions is whether the additional imaginary component of the order parameter (OP), associated with surface currents~\cite{Fogelstrom1997}, occurs only at the surface or whether it is in fact a bulk property~\cite{Dagan2002}. While a surface-associated imaginary component would still be compatible with a pure \dwave bulk OP, a bulk-associated imaginary component would not be. This distinction has important implications for our understanding of the high \Tc superconductivity mechanism. Some of the proposed models require pure bulk d-wave symmetry~\cite{Pines1994,Pines1995,Anderson1987,Anderson1988}, while others do not~\cite{Sachdev2002,Friedel2001,Sangiovanni2003}.

\par

Unlike conventional superconductors, in high-  cuprates there is much research which supports anisotropic symmetry exhibiting nodes in the order parameter, i.e. \dwave symmetry, see for instance~\cite{Dagan2000a,Tsuei2000}. In this case, the \dwave superconducting gap is estimated to be $\Delta \approx 20~meV$ for \YBCO~\cite{Dagan2000a}. Nevertheless, possible mixed states of complex order parameter ${{d}_{{{x}^{2}}-{{y}^{2}}}}+is$ or ${{d}_{{{x}^{2}}-{{y}^{2}}}}+i{{d}_{xy}}$ are also allowed within the symmetry group of \YBCO~\cite{Wenger1993}. Tunneling~\cite{Krupke1999,Sharoni2002,Dagan2000} and penetration depth~\cite{Farber2004} measurements show an additional imaginary component estimated to be of the order of $\delta \approx 2~meV$. This component was induced by magnetic field in optimally doped \YBCO films, and observed in overdoped \CaYBCO films without applying an external magnetic field.

\par

Infrared spectroscopy, which was used extensively in the field of high-\Tc cuprates, holds the same energy scale for the superconducting energy gap and the excitations energies which are assumed to be relevant for the formation of superconductivity~\cite{[{for a review see for instance }]Basov2005}. However, by using Far Infrared techniques such as THz spectroscopy, one is able to obtain the quasiparticles's dynamics far from these excitations energies spectrum. The vast use of THz spectroscopy in the past years has given us the advantage of probing bulk material properties while achieving a wide frequency range. By these means, we can focus on probing additional gap contributions in the THz range where the energy scale is of the order of 1~meV. Our work is focused on probing the complex conductivity of \CaYBCO in the THz frequency range where the energy scale of the radiation is equivalent to the energy scale of the sub-gap, $\delta \approx 2~meV$. As will be further elaborated, we tried to estimate a unique frequency and temperature dependence of the complex conductivity indicating the order parameter symmetry in our samples.

\section{Experimental}

In the current research two batches of \CaYBCO thin films were deposited. The films were grown to a thickness of 500~-~600~${\AA}$ on \LAO substrates using 10\% and 5\% Ca. The films were deposited by a DC sputtering method at a growth temperature of 820~${\circ}$C, using a pressure template procedure~\cite{Krupke2000} including additional water vapor of 2~-~3~mTorr. XRD measurements show a well oriented c-axis films. The samples' transition temperature was characterized using resistance and induction measurements. The 10\% films showed typical \Tc of 78~K$~\pm~$1~K while the 5\% Ca doped films showed typical \Tc of 84~K$~\pm~$0.7~K. From the curvature of the resistivity as a function of temperature, we have deduced that the samples are in the overdoped regime. This is done by taking the second derivative of the resistivity curve as a function of temperature.

\par

The first batch of films was measured using a Time Domain THz Spectroscopy method applying a TeraView TPS Spectra 1000\texttrademark system. Time Domain THz spectroscopy method has been reviewed in several papers~\cite{Schmuttenmaer2004,Grischkowsky2006}. The complex transmission spectra of the films, substrate and reference (clear aperture) were recorded from room temperature down to 20~K with steps of about 1~K and in the frequency range of 0.1~-~3~THz (3~-~100~\cm).

\par

The second batch of films was measured by using a THz Quasi-Optical method applying an Epsilon coherent source spectrometer. This is a well-known quasi optical method, which has been reviewed and used by several groups~\cite{[see for instance ]Gorshunov2005,*Pronin1998}. The absolute transmission function and the transmission phase shift of the films, substrate and reference were recorded from room temperature down to 20~K in steps of 50~K above 100~K and steps of 10~K below 100~K. The data was recorded for four Backward Wave Oscillators THz sources in the frequency ranges of 2~-~4~\cm (QS3-120 source), 8~-~23~\cm (QS1-370 and QS1-700 sources), and 28~-~41~\cm (QS1-1100 source).

\par

Analysis of the complex transmission in both methods was first performed for the bare substrate. The substrate exhibited unfamiliar sharp absorption at frequency of about 65~\cm due to interaction of polarization with the a-b twins which were not oriented with the sample edges. Therefore, all samples, i.e. films and substrate in both methods, were rotated by approximately 45 degrees to the polarized beam, in order to reduce this effect. In this case, we were able to measure our samples with no absorption in the entire frequency range. It is important to mention that the absorption did not affect the transmission at lower frequency range below 40~\cm. The data for the \LAO substrate was fitted using Lorentz oscillator for the phonon absorption with the resonance frequency of 190~\cm~\cite{Abrashev1999} and resulted in an almost constant dielectric function of ${{\varepsilon }_{1}}\approx 24$ in the frequency range 1 to 60~\cm at temperatures 20 to 300~K.

\par

For data analysis of our superconducting film, we used the two-fluid model. The scattering rate and plasma frequency of the quasi-particles and plasma frequency of the superconducting carriers were determined by a least-square fitting of the transmission and phase spectrum from which the complex conductivity was deduced~\cite{Berlinsky1993}. In order to achieve the best fit, we have minimized the goodness of fit parameter, ${{\chi }^{2}}$, by additionally tuning our fit parameters as a function of frequency and applying the Variable Dielectric Function~\cite{Kuzmenko2005} for the THz-TDS data and Fresnel equations~\cite{Dressel2002} for the Quasi Optical system data.

\section{Results and Discussion}

\begin{figure}
    \begin{center}
        \includegraphics[width=0.7\linewidth]{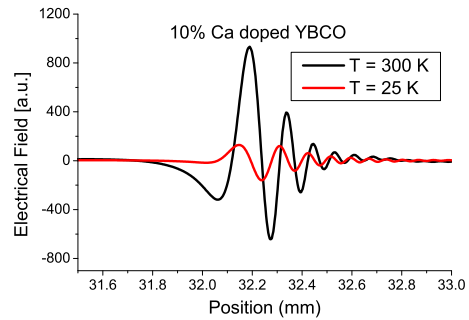}
    	\caption {Typical THz pulse measured in transmission through \CaYBCO (x=10\%) thin film at temperatures of 300~K and 25~K. Pulse exhibit reduction in amplitude and shift in position, i.e. time. Below \Tc$ \approx 78~K$ The complex transmission function is obtained by the ratio of pulse transmitted through sample to a pulse transmitted through reference.}
        \label{fig:TDS_Pulse_YBCO}
    \end{center}
\end{figure}

\begin{figure}
    \begin{center}
        \subfigure[Transmission]{\label{fig:QO_YBCO_Tr}\includegraphics[width=0.7\linewidth]{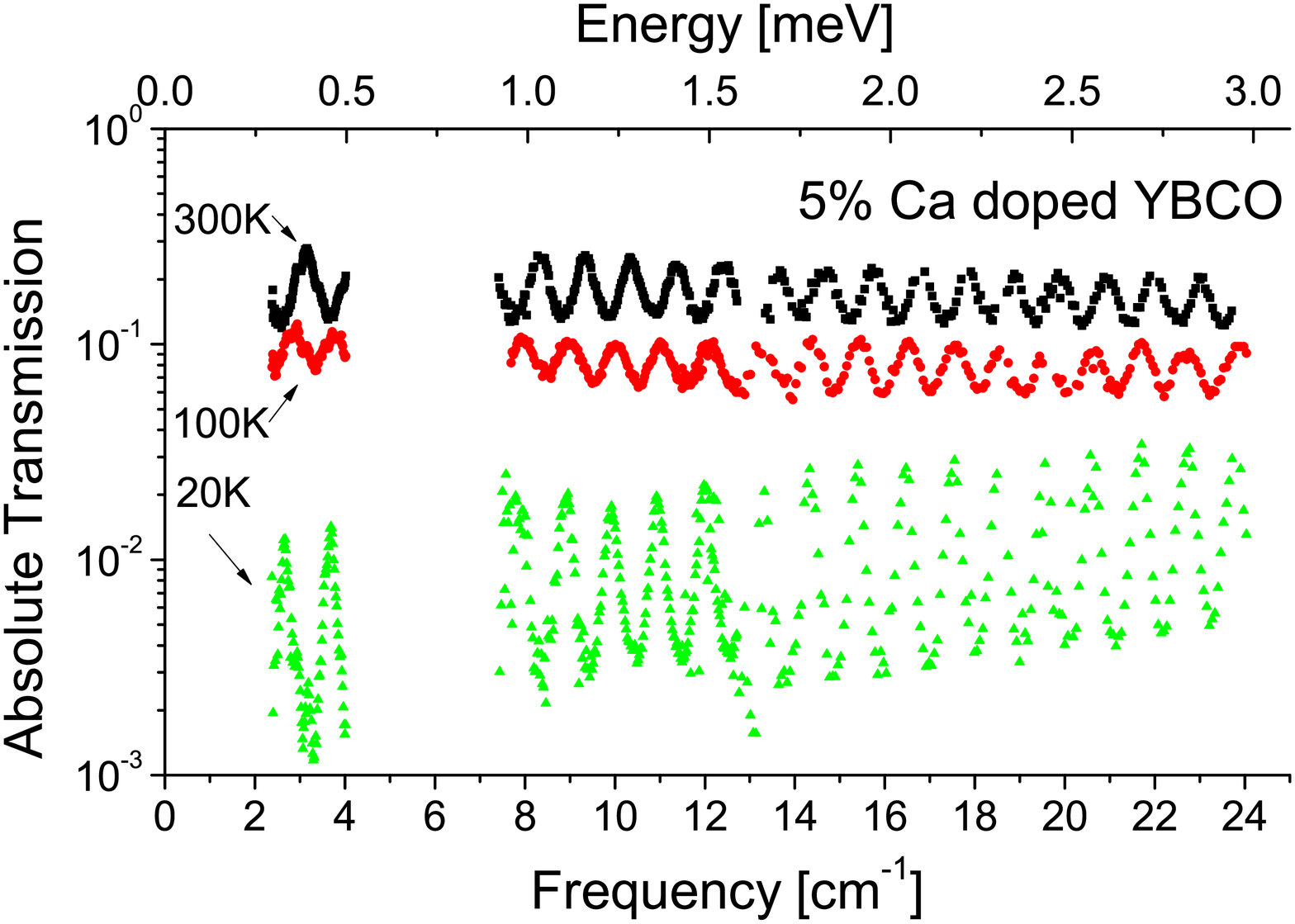}}
        \subfigure[Transmission Phase Shift]{\label{fig:QO_YBCO_Pt}\includegraphics[width=0.7\linewidth]{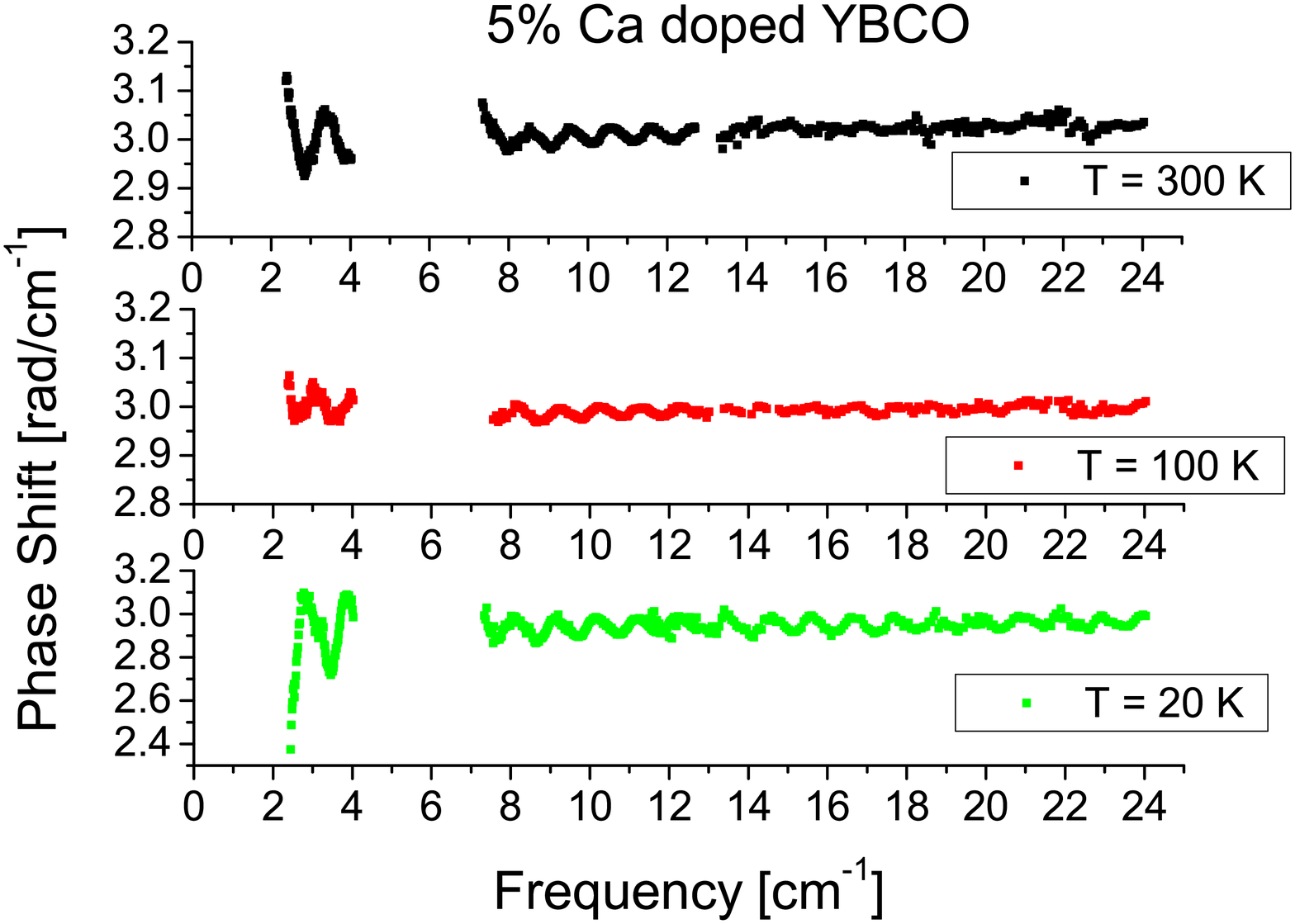}}
        \caption{The Transmission (a) and Transmission Phase Shift (b) of a \CaYBCO (x=5\%) thin film measured with the Quasi Optic system for selected temperatures and frequency ranges. Figure (a) shows decrease of transmission as cooling down and the decrease of transmission particularly at low frequencies below the critical temperature. Figure (b) shows change in phase shift amplitude and increased oscillations at low frequencies and below the critical temperature.}
        \label{fig:QO_YBCO}
    \end{center}
\end{figure}

Two batches of 5\% and 10\% \CaYBCO thin films were measured utilizing both of the spectroscopy methods. The transmission of the films decreases with temperature as shown in Figure~\ref{fig:TDS_Pulse_YBCO} and in Figure~\ref{fig:QO_YBCO_Tr} for time domain and frequency domain measurements, respectively. Below \Tc, along with a sharp decrease of transmission amplitude, there is also a change in transmission phase shift which is shown in Figure~\ref{fig:TDS_Pulse_YBCO} and in Figure~\ref{fig:QO_YBCO_Pt}. In the case of the frequency domain measurements, these drastic phase shift changes are observed mainly in the low frequency regime.

We have used the transmission function of four interfaces, i.e. sample with two layers. For the data analysis, the two-fluid model was used in order to describe the complex conductivity of the thin film
\begin{equation}
    {{\hat{\sigma }}_{1}}={{\hat{\sigma }}_{n}}\left( \omega ,T \right)+{{\hat{\sigma }}_{s}}\left( \omega ,T \right)
    \label{eqn:Two_Fluid_Conductivity}
\end{equation}
Where $\sigma_{n}$ is the complex conductivity of the normal carriers and ${{\sigma }_{s}}$ is the complex conductivity of the superconducting carriers which are described as
\begin{equation}
    \hat{\sigma}_{n} ( \omega , T ) = {{\omega}_{p}}_{n}(T) [\frac{ \Gamma(T) }{ \Gamma^{2}(T) + \omega^{2} } + i \frac{ \omega }{ \Gamma^{2}(T) + \omega^{2} } ]
    \label{eqn:Normal_Conductivity}
\end{equation}
\begin{equation}
    \hat{\sigma}_{s} ( \omega , T ) = {{\omega}_{p}}_{s}(T) [\delta( \omega =0 ) + i \frac{ 1 }{ \omega }]
    \label{eqn:Super_Conductivity}
\end{equation}
where $\Gamma(T)$ is the normal carriers' (or quasiparticles' at temperatures below \Tc) temperature dependent scattering rate, ${\omega_{p}}_{n}$ and ${\omega_{p}}_{s}$ are the plasma frequency of normal and superconducting carriers, respectively.

\par

\begin{figure}
    \begin{center}
        \subfigure[Scattering rate of the quasi particles]{\label{fig:Scattering} \includegraphics[width=0.7\linewidth]{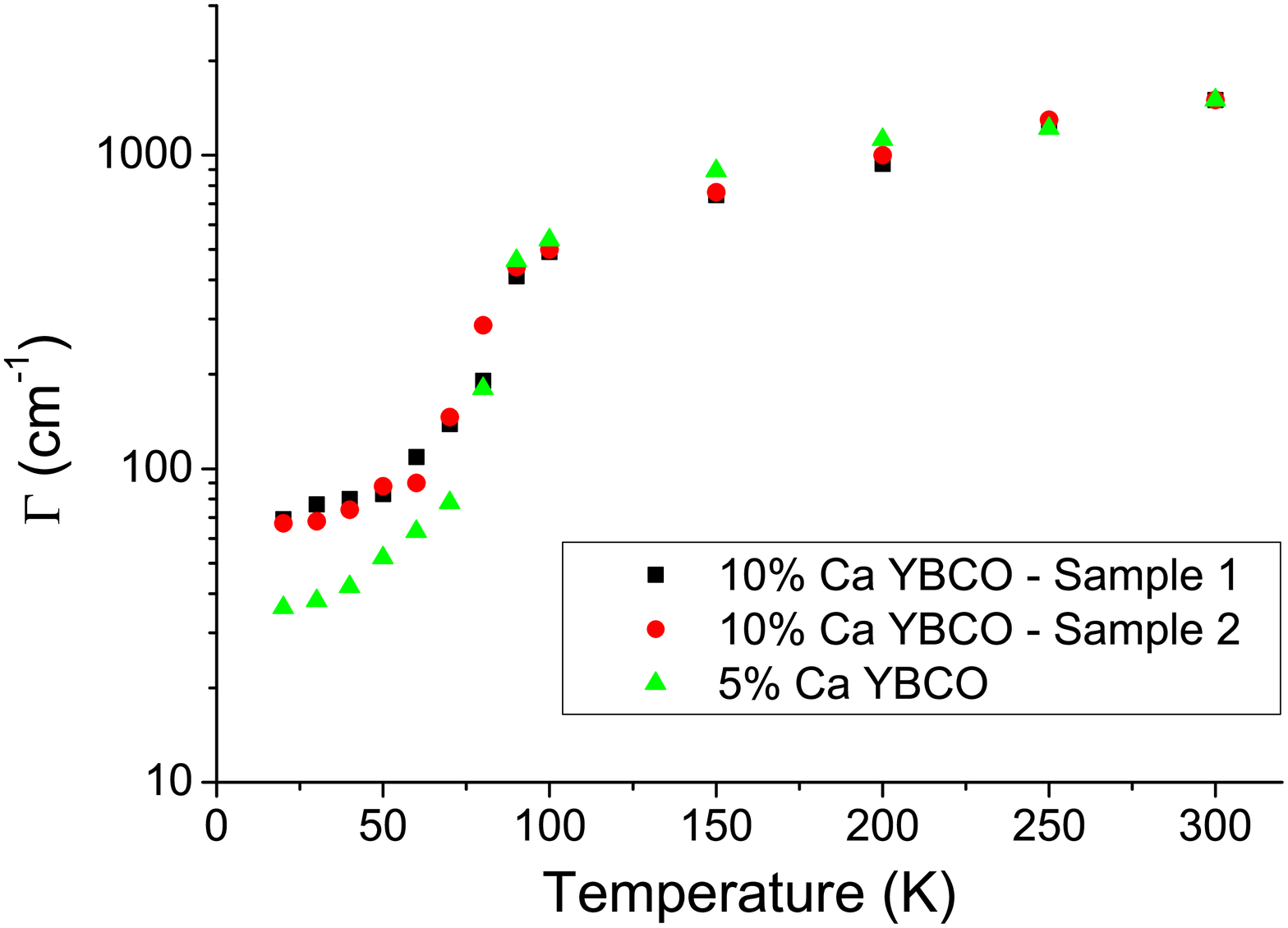}}
        \subfigure[Plasma frequency of the superconducting carriers]{\label{fig:PlasmaF} \includegraphics[width=0.7\linewidth]{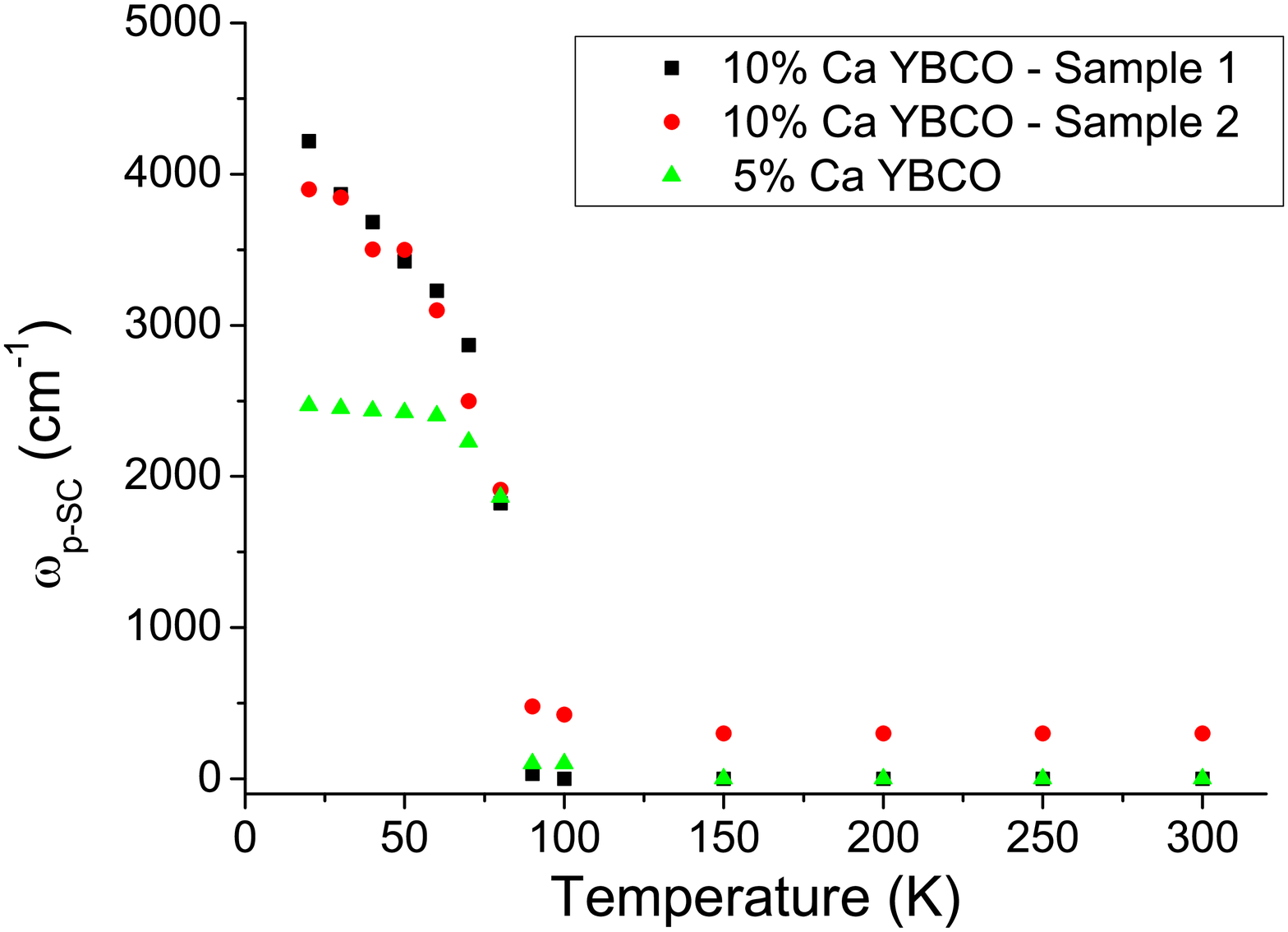}}
        \caption{Frequency independent superconducting carriers plasma frequency and quasi-particles scattering rate obtained by fitting to the two fluid model using a delta peak for the Superconducting (SC) carriers and Drude contribution for the quasi-particles. The figure shows a drastic decrease in the scattering rate at \Tc while saturating at low temperatures. At \Tc there is also an increase in $\omega_{p_{s}}$ due to the formation of pairs.}
        \label{fig:FittingParam}
    \end{center}
\end{figure}

The data analysis was confined to a change in the scattering rate of the normal carriers or quasiparticles below \Tc and change in the plasma frequencies for both types of carriers. Our analysis of the complex transmission data results in the fitting parameters' temperature dependence as shown in Figure~\ref{fig:FittingParam}. In the following section temperature and frequency dependence of the complex conductivity will be analyzed applying the frequency independent parameters of the two-fluid model followed by a frequency dependent fit. Finally, the anomaly in the complex conductivity seen at low frequencies and well below \Tc will be reviewed.

Above \Tc the temperature dependence of complex transmission is defined mainly by a decrease in its absolute value. This behavior can be understood by a decrease in the normal carriers' scattering rate in the two-fluid model as shown in Figure~\ref{fig:Scattering}. As a result, the real part of conductivity, \sigr, slightly increases as temperature decreases. This increase of \sigr above \Tc is not followed by any change in the imaginary part of conductivity, \sigi.

\par

\begin{figure}
    \begin{center}
        \subfigure[Real part of conductivity -  \sigr]{\label{fig:Sigr_QO_YBCO10} \includegraphics[width=0.7\linewidth]{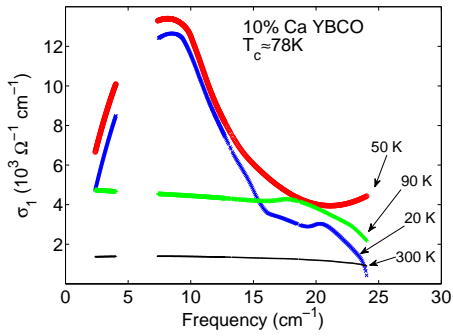}}
        \subfigure[Imaginary part of conductivity -  \sigi]{\label{fig:Sigi_QO_YBCO10} \includegraphics[width=0.7\linewidth]{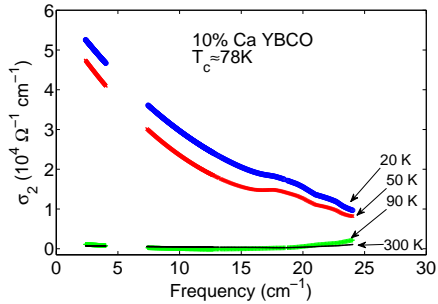}}
        \caption{The real part of conductivity as a function of frequency (b) The imaginary part of the complex conductivity as a function of frequency. Both figures represent \CaYBCO ($x=10\%$) thin film measured by the Quasi Optic method. Figure (a) shows selected temperatures above and below \Tc. Below \Tc the real part of conductivity shows a non monotonic behavior with a decrease at low frequencies. Figure (b) shows the increase of the imaginary part as $1/\omega$ according to Eq.~\ref{eqn:Super_Conductivity}.}
        \label{fig:Sig_QO_YBCO10}
    \end{center}
\end{figure}

\begin{figure}
    \begin{center}
        \subfigure[Real part of conductivity -  \sigr]{\label{fig:Sigr_TDS_YBCO10} \includegraphics[width=0.7\linewidth]{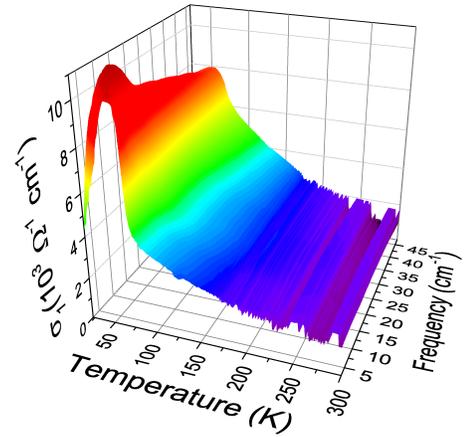}}
        \subfigure[Imaginary part of conductivity - \sigi]{\label{fig:Sigi_TDS_YBCO10} \includegraphics[width=0.7\linewidth]{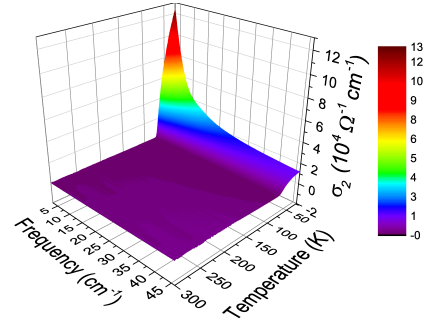}}
        \caption{The real and imaginary part of the complex conductivity for \CaYBCO ($x=10\%$) thin film measured by Time Domain Spectroscopy system. Figure (a) shows onset frequency in which the real part of the conductivity increases below \Tc, and followed by a decrease at about 10~\cm and at temperatures below 40~K. Figure (b) shows the increase of the imaginary part having $1/\omega$ dependency according to Eq.~\ref{eqn:Super_Conductivity}.}
        \label{fig:Sig_TDS_YBCO10}
    \end{center}
\end{figure}

Below \Tc there is a sharp decrease of transmission and in addition, a drastic change in the phase shift. The sharp decrease in transmission and the change in phase shift are observed mainly at low frequencies. This behavior is well explained by the formation of superconducting carriers at the transition. Therefore it is consistent with the increase in the superconducting plasma frequency as shown in Figure~\ref{fig:PlasmaF}. In addition, the scattering rate decreases sharply as shown in Figure~\ref{fig:Scattering}. This change is also related to condensation of normal carriers into the superconducting state which results in a weakening of scattering mechanisms in the film. As a result of the above changes, our analysis shows a sharp increase in both the real and imaginary parts of the conductivity as shown in Figure~\ref{fig:Sig_QO_YBCO10} and Figure~\ref{fig:Sig_TDS_YBCO10} for the frequency domain and time domain methods, respectively. It is important to note that the superconducting contribution is related mainly to the imaginary part of the conductivity (Eq.~\ref{eqn:Super_Conductivity}) in our frequency range, therefore explaining the drastic change in phase shift at low frequencies. According to Eq.~\ref{eqn:Super_Conductivity}, the imaginary part of the conductivity, \sigi, in the two-fluid model is diverging at low frequencies; this tendency is shown in Figure~\ref{fig:Sigi_QO_YBCO10} and Figure~\ref{fig:Sigi_TDS_YBCO10}. In order to understand this behavior, the change of the conductivity spectral weight below \Tc has to be considered. In the normal state and due to a high scattering rate, the real part of conductivity shows an almost frequency independent behavior in our frequency range. However, below \Tc , the scattering rate $\Gamma$ decreases drastically therefore quenching the spectral weight down to lower frequencies. The change in the spectral weight at the transition is shown in our data below \Tc where \sigr exhibit a gradual change as a function of frequency.

At lower temperatures below \Tc, our analysis shows that the scattering rate is maintaining somewhat fixed value, consistent with other reports~\cite{[see for instance ]Djordjevic2002}. In this temperature range and due to additional spectral weight removal from the Drude part of the quasi-particles to the delta peak of the superconducting carriers at zero frequency, the plasma frequency of the quasi-particles also starts to decrease. This behavior is the main cause for the "coherence-peak"-like behavior observed in the real part of the conductivity for optimally doped YBCO films~\cite{Frenkel1996} and in our current research for overdoped YBCO films~\cite{Bachar2009}.

\par

After applying the frequency independent parameters of the two-fluid model as a first approach, we can deploy other techniques in order to tune our results for achieving the best fit. In the frequency domain we have just used the Fresnel' equations~\cite{Dressel2002} for complex transmission of a two layer sample and in the time domain we have used the Variable Dielectric Function~\cite{Kuzmenko2005} for the film's data. In the frequency domain, the complex conductivity was calculated by applying the Fresnel equations on each data point without using Kramers Kroning relations. In the time domain, the VDF function is modeled as a sum of Lorentz-Drude oscillators fixed at each data point which were tuned in order to achieve the best fit. These approaches, in both techniques, enable us to obtain better agreement between our data points and a customized fit model. It is important to note that the parameters $\Gamma ,{{\omega }_{{{p}_{n}}}},{{\omega }_{{{p}_{n}}}}$ which were used in the first fit, and shown in Figure~\ref{fig:FittingParam}, were defined as frequency independent parameters. As a result, for frequencies much smaller than $\Gamma$, the real part of conductivity should be almost frequency independent. However, after applying the VDF or solving the Fresnel equations, one obtains a complex conductivity which deviates from the frequency-independent model.

\par

\begin{figure}
    \includegraphics[width=0.7\linewidth]{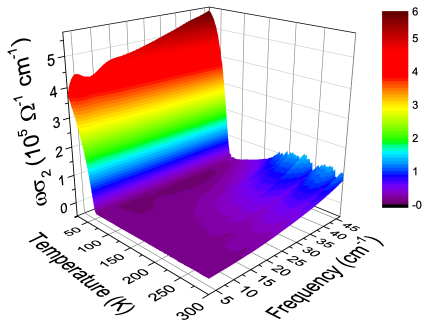}
    \caption{Imaginary part of conductivity multiplied by the frequency for 10\% Ca doped sample. The imaginary part shows a deep at frequencies equivalent to the onset in the real part}
    \label{fig:OmegaSigma2}
\end{figure}

Following the above-mentioned procedure, further analysis of the complex conductivity data below \Tc shows some known features for high-\Tc thin films. While \sigi gradually increases due to the increase in the superconducting carriers' concentration as temperature drops, \sigr, at a fixed frequency, reaches a maximum at about 40~K. This is a well-known signature for high quality thin films. The maximum conductivity as a function of temperature can also be related to the competition between changes in the scattering rate and the plasma frequency of the quasi-particles as temperature reduces. As a function of frequency, the real part of conductivity should increase or at least retain a maximum value at low frequencies. However, our data show some anomalies in the complex conductivity at low frequencies which was not reported for optimally doped \YBCO. Following this temperature and frequency dependence of \sigr, there is a sharp decrease of \sigr at low frequencies, which becomes stronger as temperature drops down to 20~K. This decrease is shown in Figure~\ref{fig:Sigr_QO_YBCO10} and Figure~\ref{fig:Sigr_TDS_YBCO10} at a frequency of about 10~\cm. The onset drop in \sigr is observed as a dip in \wsig for these overdoped samples at about the same frequency as shown in Figure~\ref{fig:OmegaSigma2}. As shown in Eq.~\ref{eqn:Super_Conductivity}, the dominating part in the imaginary \wsig is proportional to the superconducting carriers plasma frequency.

The uncommon feature observed in the real part of conductivity is inconsistent with a pure \dwave superconductor analyzed by the two fluid model. Instead of showing a constant conductivity value at low frequencies, as expected from the Drude function, the real part exhibits a drop below 10~\cm, as shown for the 10\% Ca overdoped samples. This feature appears at temperatures below 40~K and seems to be enhanced as temperature decreases.

\par

Comparing our data to previously reported data in the same frequency range using optimally doped samples~\cite{Wilke2000,Khazan2002} one can see that this non-monotonic behavior of \sigr at low frequencies and below \Tc is reported here for the first time. This behavior is observed for the first time in our current research on Ca overdoped \YBCO thin films.

\par

\begin{figure}
    \includegraphics[width=0.7\linewidth]{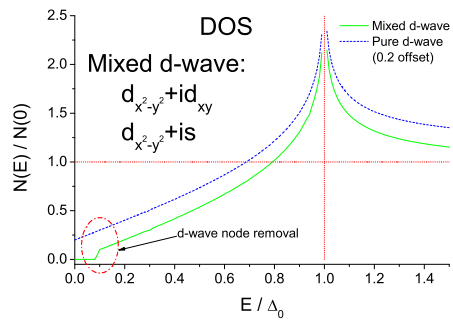}
    \caption{Simulation of the DOS as function of energy for pure \dwave and complex energy gap symmetries. The figure shows a cutoff energy scale, $\delta$, underneath no quasiparticle states are available}
    \label{fig:DOS}
\end{figure}

The current paragraph will elaborate on some possible explanations of our data. We can regard this non monotonic behavior as result of a frequency dependent scattering rate or plasma frequency which deviates, mainly at low frequencies, from the values shown in Figure~\ref{fig:FittingParam}. A decrease in the normal part of conductivity can be explained as an increase in the scattering rate or as a decrease in the quasiparticles plasma frequency at the frequency range of 10~\cm. The increase in the scattering can be compatible with Ca impurities causing localization at low frequencies. However, it is not at all clear why it should appear only at temperatures quite below \Tc. Instead, we suggest that this decrease in conductivity should be considered as a decrease in the density of states (DOS). The decrease in the plasma frequency at low frequencies is just a direct observation of this behavior. In the case of \YBCO we have compared the pure \dwave DOS to pairing symmetries such as $d_{x^{2}-y^{2}}+is$ and $d_{x^{2}-y^{2}}+id_{xy}$, i.e. mixed \dwave. Figure~\ref{fig:DOS} shows that for the pure \dwave there are states down to zero energy. However, for the complex order parameter there are no states available below a sub-gap. We suggest that this sub-gap is a result of node removal in the pure \dwave by an additional imaginary $d_{xy}$ or $s$  component. The sharp decrease in the density of states should be observed as a decrease in the quasi-particles concentration or, experimentally, by a decrease in their plasma frequency. As the real part of conductivity is a direct probe of the quasi-particles' plasma frequency, a decrease in its value is observed as this sub-gap in energy.

The idea of an additional imaginary part in a superconducting order parameter was used for calculating the complex conductivity by Sch\"{u}rrer~\etal~\cite{Schurrer1998}. In their work, $s+id$ pairing symmetry was assumed using the Eliashberg equations. In their report, several quasi-particle density of states (QDOS) spectra were calculated for different imaginary and real part ratios. In the case of small $s$ compared to the $d$ part and by assuming that the QDOS yields the absolute value of the order parameter, the results of Sch\"{u}rrer~\etal are consistent with our data. The complex conductivity which was calculated by Sch\"{u}rrer~\etal using a  $s+id$ pairing symmetry shows several striking features. The real part of conductivity, \sigr, shows an onset followed by a decrease at low frequencies which are equivalent to the sub-gap energy scale observed in our current work. In addition, the imaginary part product, \sigi, shows a dip developing at the same energy scale as the onset frequency in \sigr. The complex conductivity data in our films, which shows deviations from the pure \dwave and the frequency independent parameters of the two fluid model analysis, exhibits the same features as described above. We believe that the only possible interpretation of these deviations can be achieved by an additional imaginary part of the order parameter, as suggested by Sch\"{u}rrer~\etal.

\par

Finally, we would like to clarify at this point that the reduction of \sigr at about 10~\cm, which is assumed to be the sub-gap frequency, is at a much lower energy than the imaginary sub-gap measured directly by tunneling or obtained indirectly by microwave London penetration depth to have a value of about $\delta \approx 2~meV (\approx 16~cm^{-1})$~\cite{Farber2004}. We cannot rule out a possible explanation for this discrepancy where this extra imaginary gap component is a surface effect and reduces once bulk is examined.

\section{Conclusions}

The complex conductivity of Ca doped \YBCO thin films was measured by both frequency and time domain spectroscopy techniques in the THz frequency range. We have shown distinct evidence for a sub energy gap in the THz frequency range which develops below \Tc and exhibits itself as a significant decrease in the real part of conductivity with a typical onset frequency of 10~\cm. The sub-gap in the real part of the conductivity shows up as a dip in \wsig at the same frequency, representing a decrease in the superconducting plasma frequency. These features in the measured complex conductivity spectrum match the calculated spectrum by Sch\"{u}rrer~\etal using a complex order parameter with imaginary part of $2\delta \approx 1.2~meV$. We suggest that this gap is related to node removal in the overdoped regime of \YBCO. This idea has previously been reported utilizing different surface sensitive tunneling experiments and microwave indirect measurements. However, this gap appears at lower energies than in previous research. At the current stage we cannot rule out the effect of Ca atoms, which might act as impurity centers causing localization and this could be done by a future research in which the Ca doped samples are annealed into the optimal doped regime. In order to avoid the influence of a-b twins on frequency response, caused by the substrate, which is presumably temperature independent, overdoped thin films, should be grown on un-twinned substrates.

%et al. example
%\etal\cite{Rennerandfisher}

%example
%\textit{d}-wave

\begin{acknowledgments}

The authors would like to thank D. Van Der Marel for using his TeraView system. We want to thank G. Deutscher for useful discussions and for using his sputtering system. This work was supported in part by the Ministry of Science and Technology (MOST), State of Israel, and by the RAS "Program for fundamental research Problems of Radiophysics".

\end{acknowledgments}

\bibliographystyle{apsrev4-1}
\bibliography{Bachar_arXiv_Bib}

\end{document}